\documentstyle[11pt,epsfig]{article}
\newcommand \be{\begin{equation}}
\newcommand \bea{\begin{eqnarray}}
\newcommand \ee{\end{equation}}
\newcommand \eea{\end{eqnarray}}

\begin{document}

\title{Could short selling make financial markets tumble?}

\author{J\o rgen Vitting Andersen$^{1,2}$\\
$^1$ U. F. R. de Sciences \'Economiques, Gestion, Math\'ematiques et
Informatique, \\ CNRS UMR7536 and Universit\'e Paris X-Nanterre,
92001 Nanterre Cedex, France \\
$^2$ Laboratoire de Physique de la Mati\`{e}re Condens\'{e}e\\ CNRS UMR6622 and
Universit\'{e} de Nice-Sophia Antipolis, 06108 Nice Cedex 2, France\\
e-mail: vitting@unice.fr\\}

\date{\today}
\maketitle

{\bf
It is suggested to consider long term trends of financial markets 
as a growth phenomenon. The question that is asked is what conditions are 
needed for a long term sustainable growth or contraction in a financial market? 
The paper discuss the role of traditional market players 
of long only mutual funds versus 
hedge funds which take both short and long positions.
It will be argued that financial 
markets since their very origin and only till very recently, have been 
in a state of ``broken symmetry'' which favored long term growth instead of 
contraction. The reason for this ``broken symmetry'' 
into a long term ``bull phase'' is the historical 
almost complete dominance by long only players in financial markets. 
Dangers connected
to short trading are illustrated by the appearence of long term
bearish trends seen in analytical results and by
 simulation results of an agent based market model. 
Recent short trade data of the Nasdaq Composite index show an increase in the
short activity prior to
or at the same time as dips in the market, and reveal 
an steadily increase in the short trading activity, reaching 
levels never seen before.
}

\vskip 0.5cm
\vskip 0.5cm
Stock markets are known to be notoriously difficult to predict. 
The impossibility of predicting has been formalized in the Efficient 
Market Hypothesis (EMH) which 
states that all available information is included in
prices\cite{Fama,Samuelson}. Still without attempting specific prediction, it 
makes very well sense to ask which conditions are needed for a 
substainable trend, e.g. what is the ``fuel'' needed for a manifestation 
of a longer term bull or bear market? 
In this paper it is suggested to consider long term trends of financial markets 
as a growth phenomenon. 
The paper will discuss the role of traditional market players 
of long only mutual funds versus 
the recent arrival of hedge funds which take both short and long positions.
Short selling\cite{Chen,Platt,Ackert}  happens when speculators sell,  
 say at time $t_0$ to price $P(t_0)$, securities
that they do not 
own but borrow. At a later time, say $t_1$, the speculator has to 
buy the security and give it back to it's owner. 
In case of a price decline between 
$t_0$ and $t_1$ the speculator thereby makes the profit $P(t_0) -P(t_1)$. 
This is in contrast to long only funds who are only allowed to own 
securities and therefore can only profit when prices rise. 
To make an analogy with a physical system, it will be argued that financial 
markets since their very origin and only till very recently, have been 
in a state of ``broken symmetry'' which favored long term growth instead of 
contraction. The reason for this ``broken symmetry'' 
into a long term ``bull phase'' is the historical 
almost complete dominance by long only players in financial markets. Only 
with the recent arrival of investors that take up short positions is the 
symmetry slowly being restored, with the implications, as will be argued  
in this paper, of an increased probability for lasting decline of the markets, 
i.e. appearance of a long term ``bear phase''. 

Let us for a moment imagine a financial market could be completely 
determined by the action of just one powerful investor.
This is in contrast to traditional 
economic theory which view the stock market as reflecting the state of economy, 
i.e. the prices of stocks should reflect present earnings as well as 
the expectation value of all future earnings of a given company. An 
alternative view would be to consider the stock market as the ``motor'' 
of the economy illustrated e.g. by the so called ``wealth effect'', where 
increasing stock market prices leads to higher consumer confidence and 
spending, which in turns leads to larger earnings of companies making 
stocks rise, etc. ad infinitum. With this in mind, one could 
ask how big a fortune 
would such an investor need and which expectations 
about future developments in 
interest rate and dividends would ensure  
the investor to maintain a given growth of the market to his own profit?
In order to answer these questions, consider first the case of a large 
mutual long-only fund that tries to control the market. The wealth of 
the fund, $W$, at any given time $t$ is:  
\be
W(t) = n(t) P(t) + C(t) ~,   \label{wealth}
\ee
where $P(t)$ is the price of the market, $n(t)$ is the number of market shares held 
and $C(t)$ the cash possessed by the fund at time $t$\cite{note1}. 
At each time step the dominating fund buys new market shares in 
order to push up the price of the market, creating an excess 
demand of market shares, $A(t)$.
$A(t)$, give rise to the following equation 
for the return $r(t)$ of the market \cite{bouchaudcont,farmer}: 
\be
r(t) \equiv \ln (P(t+1)) - \ln (P(t)) =  A(t)/\lambda ,  \label{price-eq}
\ee
where $P(t)$ is the price of the market 
and $\lambda$ is the liquidity of the market. The fact that the price
goes in the direction of the sign of the order imbalance $A(t)$ is 
well-documented
\cite{Holthausen,Lakonishok,Chan,Maslov,ChalletStin,Plerou}.
$A(t)$ will first be taken a constant, $A(t) \equiv A$ i.e. we have 
\be
r(t) \approx {d \ln{P(t)} \over dt} = A / \lambda ; \ \  P(t) = e^{A t/\lambda}, \label{P}
\ee
and 
\be
{d n(t) \over dt} = A ;   \ \ n(t) = A t, \label{n}
\ee
Besides expenses to keep on buying shares, the mutual fund get an income 
from dividends, $d(t)$,  of the shares it's already holding, and from 
interest rates, $r(t)$, of its cash supply $C(t)$. As will be seen the 
expectation of future dividends $d(t)$ and future interest rates $r(t)$ is 
crucial in designing a profitable strategy for the mutual fund. Since 
the aim of this paper is to study long term trends where changes in 
dividends can be much more important compared to changes in interest rates,
$r(t)$ will be taken constant, $r(t) \equiv r$, and the focus will be the 
time dependence of $d(t)$. Consider first the simplest case where also 
$d$ is constant: $d(t) \equiv d$. The balance equation for the cash supply 
as a function of time therefore reads\cite{note5}:  
\bea
{d C(t) \over dt} &  =  &  -{d n \over dt}  P(t) + C(t) r + n(t)  d  
+ C_{\rm
flow}(t,r(t),d(t),P(t),...)\\ 
&  = & - A  e^{A t/\lambda} + C(t) r +A t d 
+ C_{\rm
flow}(t,r(t),d(t),P(t),...)
\label{dC}
\eea
The term $ C_{\rm flow}(t,r(t),d(t),P(t),...)$ describes the possibility of 
additional inflow/outflow  of 
money into the fund and as indicated can be thought of to depend on time, 
dividends, interest rates, the price of the market 
as well as many other factors like e.g. tax cuts etc\cite{Sornette1}.  
In practice the inflow/outflow of liquidity often  contributes a 
significant part to the total wealth of a mutual fund, but for 
clarity $ C_{\rm flow}(t,r(t),d(t),P(t)...)$ will be taken equal to 0 first, 
before discussing the relevance of various functional forms below.
It is preferable to express (\ref{dC}) in terms of the growth rate of 
the financial market, $\alpha \equiv A/\lambda$  and the cash in 
terms of the market liquidity, $\bar{C} \equiv C/ \lambda$ in which 
case (\ref{dC}) become:  

\be
{d \bar{C}(t) \over dt} 
 = - \alpha  e^{\alpha t} + \bar{C}(t) r + \alpha t d \label{dC1}
\ee
The solution of (\ref{dC1}) reads:
\be
 \bar{C}(t) =  {\alpha e^{\alpha t} \over r - \alpha} 
  - {\alpha d t \over r} 
  - {\alpha d  \over r^2} 
 +  { e^{r t} \over r - \alpha} \{- \alpha -{\alpha^2 d \over r^2}
 +{\alpha d  \over r}
 + C_0 (r - \alpha) \} \label{C1}
\ee
The condition to determine whether the mutual fund can impose 
a given growth of the stock market is that $C(t) > P(t) \ \ \forall t$ 
, i.e. the cash supply should always be large enough that yet another 
market share can be bought. 
The time dependence of (\ref{C1}) is dominated by the 
factors $e^{\alpha t}$ and $e^{r t}$ expressing the growth 
due to the stock market and interest rates respectively. For 
a stock market with a growth rate larger than the interest 
rate, $\alpha > r$ the first term on the r.h.s. of (\ref{C1}) 
dominates and becomes negative. In this case the growth of the 
stock market is not sustainable, and depending on the values 
of $r,d, C_0$ the investor will inevitably face a critical time, $t_c$,
without enough money to secure further growth of the stock 
market. This situation is illustrated by the liquidity curve b) in Fig.~1 for 
the case $\alpha = 0.2, r=0.1$ and $d=0.02$ \cite{note6}: 
after approximately 750 
days the liquidity of the investor falls below the price of the 
stock market given by curve a) and he can no longer secure the 20\% 
growth of the stock market. Had the investor received four times 
more in dividend, illustrated by curve c), $t_c$ would only be delayed
by some few tens of days. Increasing the initial amount of liquidity, curve 
d), push $t_c$ further to the future, but eventually,  
for a ``super'' interest 
rate growth of the stock market, the investor will fall short 
of money and the growth have to stop. On the other hand for 
``sub'' interest rate growth $\alpha < r$ the term $e^{r t}$ 
dominates and a sustainable growth depends on the values of 
$(\alpha,r,d)$. A $\alpha=7 \%$ growth, curve e), with interest rates 
at $r= 10 \%$ and dividends $d= 2\%$ can be sustained if the initial amount of money is 
large enough curve f) whereas lowering the initial liquidity the 
growth no longer becomes sustainable, curve g).  

The Eq.~(\ref{C1}) illustrates the case that the expectation of a 
constant dividend $d$ is not sufficient for a super interest 
rate market growth as e.g. was seen in the booming stock market growth 
of the ninetees.  Clearly in that case the rise of the stock market 
was closely related to the sky rocketing expectations (justified or not) 
of future dividends. Using the Efficient Market Hypothesis  an 
investor would expect 
the dividend to follow the growth of the stock market, 
and as will be seen super 
interest rate growth become possible in that case. 
If one take the dividend to increase proportional to the  
price $d(t) \equiv d_0 P(t)$  (\ref{dC1}) takes the form:
\be
{d \bar{C}(t) \over dt} 
 =  - \alpha  e^{\alpha t} + \bar{C}(t) r 
+ \alpha t e^{\alpha t} d_0 \label{dC_dP}
\ee
with the solution: 

\be
 \bar{C}(t) =  \alpha e^{\alpha t} \{ 
   {t d_0 - 1 \over \alpha - r} 
  - {d_0  \over (\alpha - r )^2}  \}
 +   e^{r t} \{ {-r \alpha + \alpha^2  + \alpha d_0  
\over (\alpha - r )^2} 
 + C_0 \} \label{C_dP}
\ee
Curve b) in Fig.~2 illustrates a solution of (\ref{C_dP}) where a given
amount of initial liquidity and a  interest rate $r$ of  10\%,
 a dividend $d_0$ of 2\% is not enough to sustain a stock market 
growth of 20\% (curve a)). On the other hand, for a higher initial 
dividend $d_0$ of 8\% and same initial amount of liquidity, super 
interest growth of 20\% does become sustainable (i.e. $C(t) \geq P(t)
  \ \ \forall
t$) as illustrated 
by curve c).  Sufficient initial funds are however required as illustrated 
by curve d) where a high initial dividend of 8\% is not sufficient 
to avoid that the investor runs out of money ($C(t) < P(t)$  for $t \approx 
2800$). For each set of values $(d_0,C_0)$ there exists a range of 
growth potentials, $\alpha$,  for the investor $r < \alpha_{\rm low}
< \alpha < \alpha_{\rm high}$. If the investor chooses to push up the 
market at a too slow growth rate $\alpha < \alpha_{\rm low}$ he eventually 
will run out of money at large time scales, whereas for a too fast growth $\alpha > 
\alpha_{\rm high}$ the investor will miss money to continue buying shares at 
short time scales. For the case illustrated by curve c) in Fig.~2, $\alpha_{\rm
low}=0.18, \alpha_{\rm high}=0.32$.

The discussion so far has been on a large mutual long-only fund that 
tries to benefit from a constant growth in a stock market. Eqs.(\ref{dC}) and 
(\ref{dC_dP}) also describe the more morbid situation where a hedge fund 
tries to benefit from a constant decline in a stock market by short selling.
This situation 
is described in (\ref{dC_dP}) 
taking $\alpha$ and $d_0$ negative. $d_0$ should be taken negative since a 
short seller has to pay dividend to the owner from which the stock 
was borrowed. Instead of the condition $C(t) > P(t)$ that was needed 
at all times for a mutual fund to keep on buying shares, a hedge fund 
would need to ensure that $W(t) > |n| P(t)$, i.e. the wealth of the 
hedge fund has to be sufficient that it can at any time buy back 
all the shares $n$ that it has borrowed (and sold) at an earlier 
time\cite{note2}. Figure~3 illustrates the  case c) of Figure~2 
with an interest rate of $r=10 \%$ and dividend $d_0= 8\%$ and the 
same initial liquidity $C_0=10$ but now $\alpha=-20\%$ (curve a). 
Since the wealth $W(t)$, curve c), is above $|n| P(t)$, curve b), it 
is indeed sustainable (and very profitable (!) as seen from the wealth 
curve)  for a dominating hedge fund to force a continuous decline of 
the stock market. Notice that the case of a dominating long-only 
mutual fund that pushes the prices up versus the case of a dominating 
short-only hedge fund that makes the market tumble is only 
symmetric on short time scales. 
This is illustrated in curve d) in Fig.~3 which shows the 
wealth of a long-only fund with same parameters as the short-only 
fund curve c), except for the signs of $\alpha$ and $d$. Since 
the difference is however small on time scales relevant for 
trading (e.g. the difference in wealth is less than 2\% after 
5 years for the example given in Fig.~3) the difference will 
be ignored in the following.

One can  think of many different and relevant expressions of the 
flow term $C_{\rm flow}(t,r(t),d(t),P(t),...)$. The simplest case 
where customers add/subtract a constant flow of money per time unit, 
$C_{\rm flow}(t,r(t),d(t),P(t),...)= \mu$, would not change anything 
of the analysis presented of (\ref{dC}) and (\ref{dC_dP}) since 
$e^{\alpha t}$ and $e^{r t}$ would still be the dominant factors. 
Taking into account the interest rate in the inflow/outflow, 
$C_{\rm flow}(t,r(t),d(t),P(t),...)= \mu e^{r t}$, would 
create a dominant $\mu t e^{r t}$ term in the case of a constant 
dividend (\ref{dC}). However in the more relevant case where the 
dividend follows the price, $d =d_0 P(t)$ (\ref{dC_dP}) super 
interest market growth would still be dominated by the 
${t d_0 \over \alpha - r}\alpha e^{\alpha t}$ term and 
the conclusions presented above would remain 
unchanged. \ \\

The eqs. (\ref{dC}) and (\ref{dC_dP}) express the case of one single 
powerful investor that has enough liquidity to drive the market. This 
is clearly not a very realistic situation, but as will be seen it helps 
understanding the much more realistic and complex case of 
a group of investors that all try to drive the market for their own 
profit. One very much studied market model with competing agents is 
the so called ``Minority Game'' (MG) introduced by Challet and Zhang 
\cite{minoritygames-standard}. In the ``Minority Game'' the  
first approximation is to include only the direction of 
a market move by representing a financial time series as  
a binary time series $B(t)$ with 0 corresponding to a down 
move and 1 to an up move. 
Having in total $N$ agents, each agent $i$ possesses a 
memory of the last $m$ digits of $B(t)$.
A strategy gives a prediction for the next
outcome of $B(t)$ based on the history of the last $m$
digits of $B$. Since there are $2^m$ possible histories, the total number
of strategies is given by $S=2^{2^m}$. Each agent holds the same number $s$ 
(in general different) of strategies among the $S$ possible strategies. 
At each time $t$, every agent uses her most successful strategy (in
terms of payoff, see below) to decide whether to buy or sell
an asset. The agent takes
an action $a_i(t) = \pm 1$ where $1$ is interpreted as
buying an asset and $-1$ as selling an asset. The excess demand, 
$A(t)$, at time $t$ is
therefore given as $A(t) = \sum_{i=1}^{N} a_i(t)$. 
The return and the price of the market is as before given by (\ref{price-eq})
In the MG the payoff function of a strategy, $g^{\rm MG}_i(t)$, 
was rewarding strategies that a each time step 
belong to the minority decision: 
$ g^{\rm MG}_i(t) =  -a_i(t) A(t)$. 
However as stressed in another paper, 
(\cite{Vitting}), 
in real markets, the driving force underlying the competition between investors 
is not a struggle to be in the minority at each time step, but rather a fierce 
competition to gain money. Introducing profit as the selection mechanism for
the strategies the payoff function was shown to take the form\cite{Vitting}: 
$g^\$_i(t+1) =  a_i(t) A(t+1)$
The same payoff function was also used in another market model by Giardina and 
Bouchaud\cite{Giardina1,Giardina2}.
Consequently the market model with this payoff function was called the 
$\$$-game as a reminder that profit drives the selection of strategies. 

Figure~4 illustrates one price history (fat solid line) of the $\$$-game
with $N=20$ agents, $r=10\%$, $d_0=8\%$ assuming like before that the dividend
follows the market price $d(t)=d_0 P(t)$. The parameter values used were $s=4, 
m=8, C_0=50$. Stochasticity was introduced via the initial strategies hold 
by the agents as well as by $N$ 
additional ``noise'' agents 
who 
at each time step made a random decision to either sell or buy one 
market share. The liquidity parameter, $\lambda$, in  (\ref{price-eq}) 
was determined using the assumption that an univocal decision of 
all the agents to sell an asset should lead to a ``crash'' of 10\%, i.e. 
from the condition $r(t)=-2N/\lambda=-10\%$. The 10\% was chosen to 
mimic some of the very largest crashes seen in financial markets. 
Thin dotted lines represent the 5\%, 
50\% and 95\% quantiles (from bottom to top) respectively, i.e. at every 
time $t$ out of 
the 1000 different initial configurations only 50 got below the 5\% 
quantile line, 500 got below the 50\% quantile and 950 below the 95
\% quantile. The {\em average} behavior of an 
agent\footnote {when using the word ``agent'' 
hereafter is meant to refer to the $N$ agents that use strategies in 
the decision to buy or sell} 
can now be 
understood using the analysis of eq.'s (\ref{wealth})-(\ref{C_dP})

All the 
$N$ agents in figure~4 were long-only, i.e. they correspond to the 
case represented in figure~2 curve c). As predicted from the analysis 
of (\ref{dC_dP}) the price $P(t)$ is {\em in average} 
tilted towards positive returns. 
Figure~5 represents simulations with the same parameter values as in 
figure~4 except that a fraction, $\rho =0.20$ of the agents can be 
both short and long. As can be seen from the 5\% quantile, the 
introduction of agents that can take short positions clearly 
increases the probability significantly for a lasting bearish 
trend. Increasing $\rho$ to 0.4 amplifies this tendency, again 
in accordance with (\ref{dC_dP}) as illustrated by the short-only 
hedge fund figure~3 curve c. \ \\ 

Figure~7 shows the Nasdaq Composite index (fat solid line) as well as 
statistics of short selling over the time period 
1/1 2000 - 1/8 2003. Dotted line represents the fraction 
of change in shares shorted, $n_s(t)$, defined 
as the monthly total number of 
change in shares shorted divided by the total monthly 
share volume\cite{note4}. The dashed line represents the cumulative fraction 
of change in shares shorted, $Q_s(t) = \sum_{t'=0}^t n_s(t')$. 
Both $n_s(t)$ and $Q_s(t)$ have been multiplied with a factor 
$10^5$ in order to make possible a comparison with the temporal 
evolution of the Nasdaq index. 
An interesting quantity in the present context 
would be the short activity, $A_s(t)$, simply defined as the 
total volume of short transactions divided by the total volume 
of transactions. Notice that $n_s(t)$ gives the lower bound 
on $A_s(t)$ since short positions opened and closed (or vice versa) 
within a month are not counted in $n_s(t)$ whereas they would 
be in $A_s(t)$. Since $n_s(t)$ typically varies 
between 0.1-0.6 \% it is not unreasonable to assume $A_s(t)$ to 
be one order of magnitude larger, i.e. some few \%. This means 
that the short activity on the Nasdaq index could be in the 
neighborhood that lead to a high probability of a lasting 
bearish trend seen in e.g. Fig.~4-6. 

Simple visual inspection 
of the short trade data in Fig.~7 reveals an increase in the 
short activity $n_s(t)$ prior to 
or at the same time as dips in the market 
(see e.g. spring and fall 2000, summer 2001, fall 2002). 
Looking at the overall activity clearly confirms this pattern 
with a steadily increase in the cumulative short trading activity 
$Q_s(t)$ over the time period 2000 till summer 2001 and a 
steadily decrease of the market in the same time period. It is 
puzzling however,  that one see an 
continuous increase of short activity so far in 2003 
where the market also has made an increase. 
{\em Either} the market is right and short sellers will have to
cover their positions in further progress of the market (i.e. 
$Q_s(t)$ should decrease in the future), {\em or} the 
short sellers got it right and the market is in for a future
dip\cite{Sornette2}.  \ \\

It is clearly not easy to estimate precisely what impact 
short trading has on the evolution of financial markets. Still 
this paper has tried to point out some of the dangers connected 
to short trading as illustrated by the appearing of long term 
bearish trends seen in the analysis of (\ref{dC_dP}) and by 
the simulation results in Fig.~4-6. It should be noted from 
Fig.~4-6 that even as the fraction of agents that use short 
trading increase, the probability for a long term ``bullish'' 
phase remain unchanged as illustrated by the 95\% quantile lines 
which are unchanged in all three cases. This highlights  
that the danger with short selling is not when the markets are 
already in a long term bullish phase, since short sellers then 
are forced to cover their positions. The real danger is instead 
when a downwards spiral of the markets has begun, in which case 
an increase in short trading activity will only increase the 
downward trend. Looking at the bad performance of the markets 
over the past years together with the increase in short selling 
activity, which was virtually absent just few years ago, the present 
study suggest the possibility that financial markets could have entered 
a new era not seen so far. \ \\

The author thanks D. Sornette for useful discussions.

\begin{figure}[h]
\includegraphics[width=16cm]{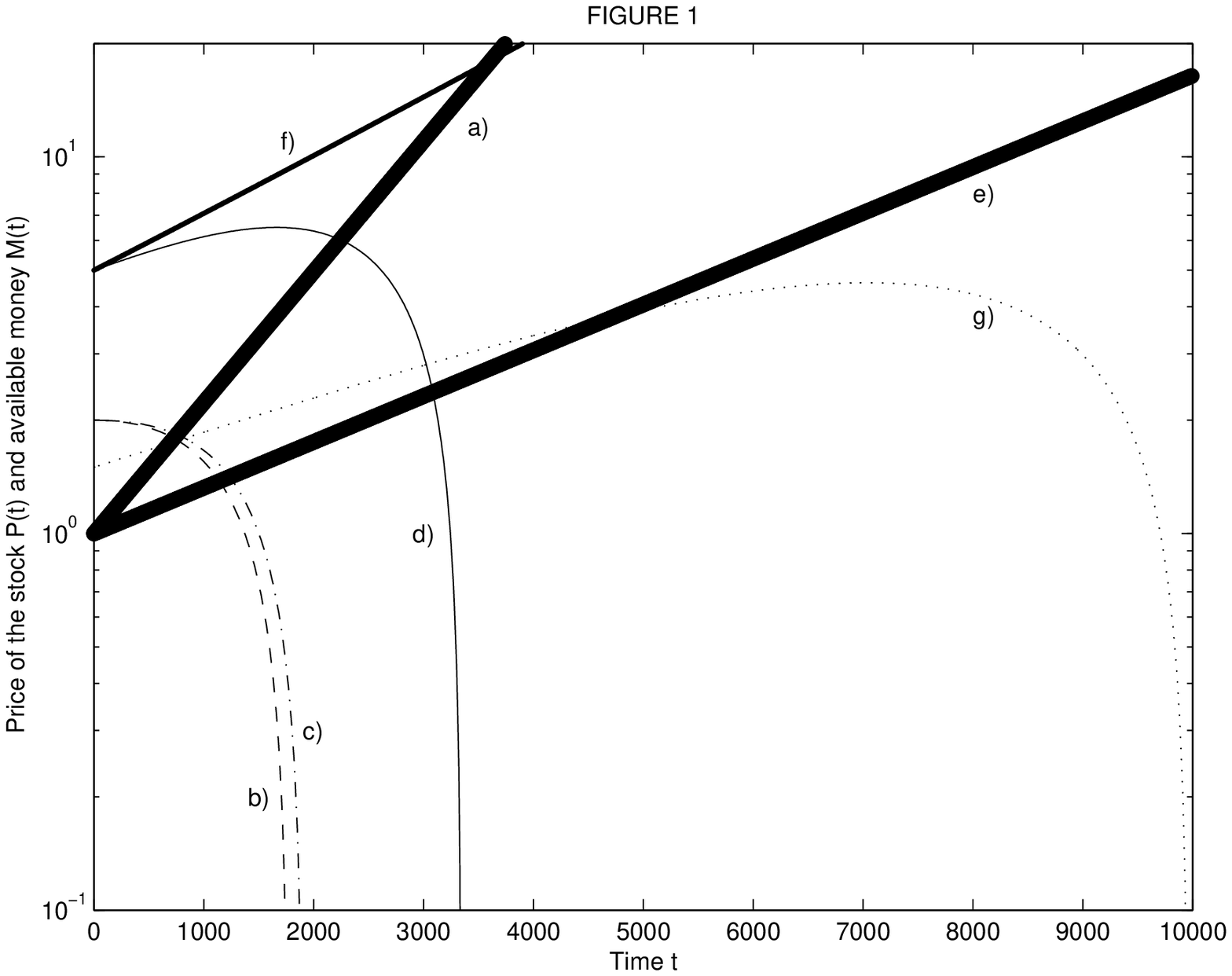}
\caption{\protect\label{Fig1} 
a) Price $P(t)$ of (\ref{P}) as a function of time $t$ for 
$\alpha = 0.2$.
 b) Cash $C(t)$ solution of (\ref{dC}) for 
the case $\alpha = 0.2, r=0.1, C_0=2$ and $d=0.02$. 
c) Similar to curve b) but with $d=0.08$. d) similar to c) but 
with $C_0=5$.
e) Price $P(t)$ of (\ref{P}) for 
$\alpha = 0.07$.
f) Cash $C(t)$ solution of (\ref{dC}) for 
the case $\alpha = 0.07, r=0.1, C_0=5$ and $d=0.02$. 
c) Similar f) but with $C_0=2$.
}
\end{figure}

\begin{figure}[h]
\includegraphics[width=16cm]{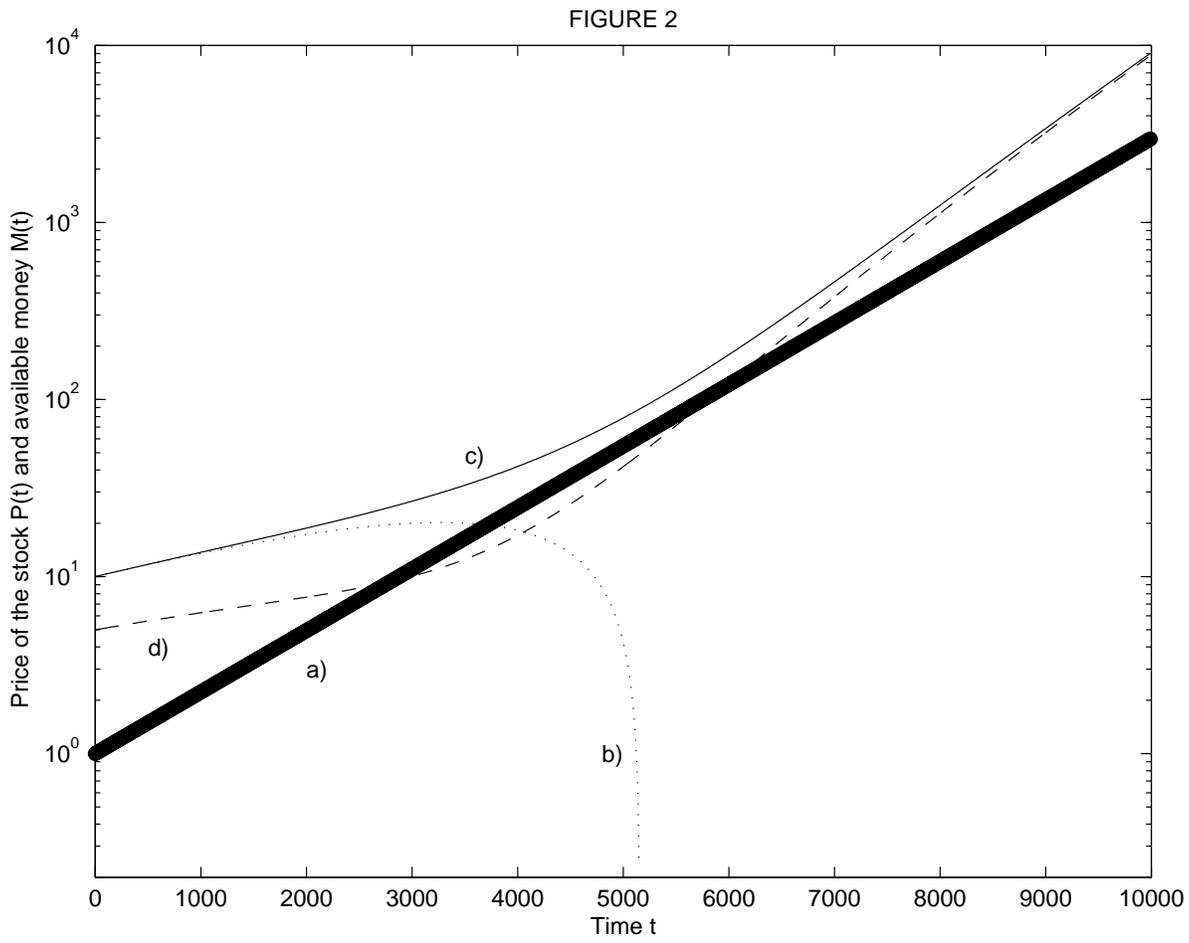}
\caption{\protect\label{Fig2} 
a) Price $P(t)$ of (\ref{P}) as a function of time $t$ for 
$\alpha = 0.2$.
 b) Cash $C(t)$ solution of (\ref{dC_dP}) for 
the case $\alpha = 0.2, r=0.1, C_0=10$ and $d=0.02$. 
c) Similar to curve b) but with $d=0.08$. d) similar to c) but 
with $C_0=5$.
}
\end{figure}

\begin{figure}[h]
\includegraphics[width=16cm]{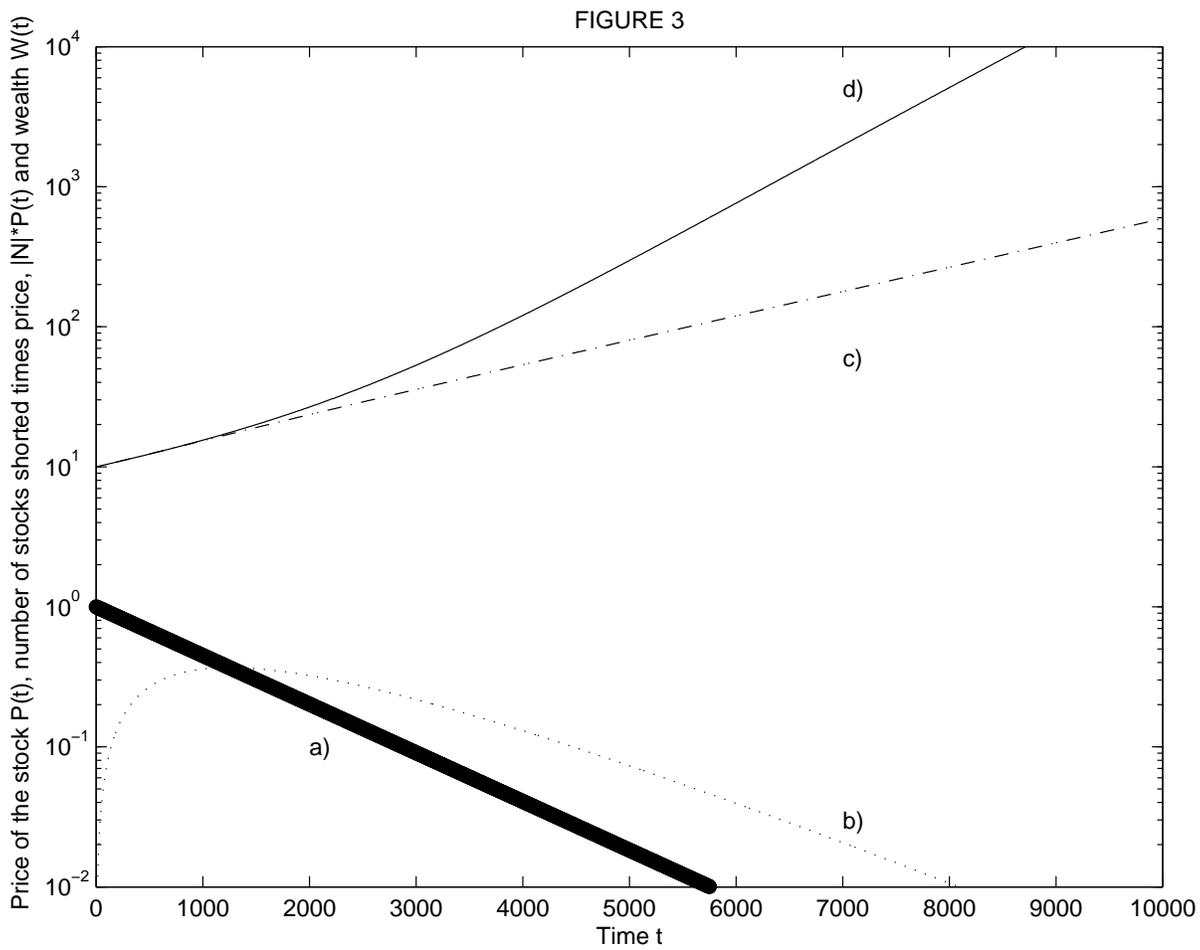}
\caption{\protect\label{Fig3}
a) Price $P(t)$ of (\ref{P}) as a function of time $t$ for 
$\alpha = -0.2$.
 b) $|n|P(t)$ for 
the case $\alpha = -0.2, r=0.1, C_0=10$ and $d=-0.08$. 
c) Wealth, $W(t)$ as a solution of (\ref{dC_dP}) and (\ref{wealth}) 
for the parameters of curve b). Same as curve b) except 
for signs of $\alpha, d$, $\alpha = 0.2$ and $d=0.08$.
}
\end{figure}

\begin{figure}[h]
\includegraphics[width=16cm]{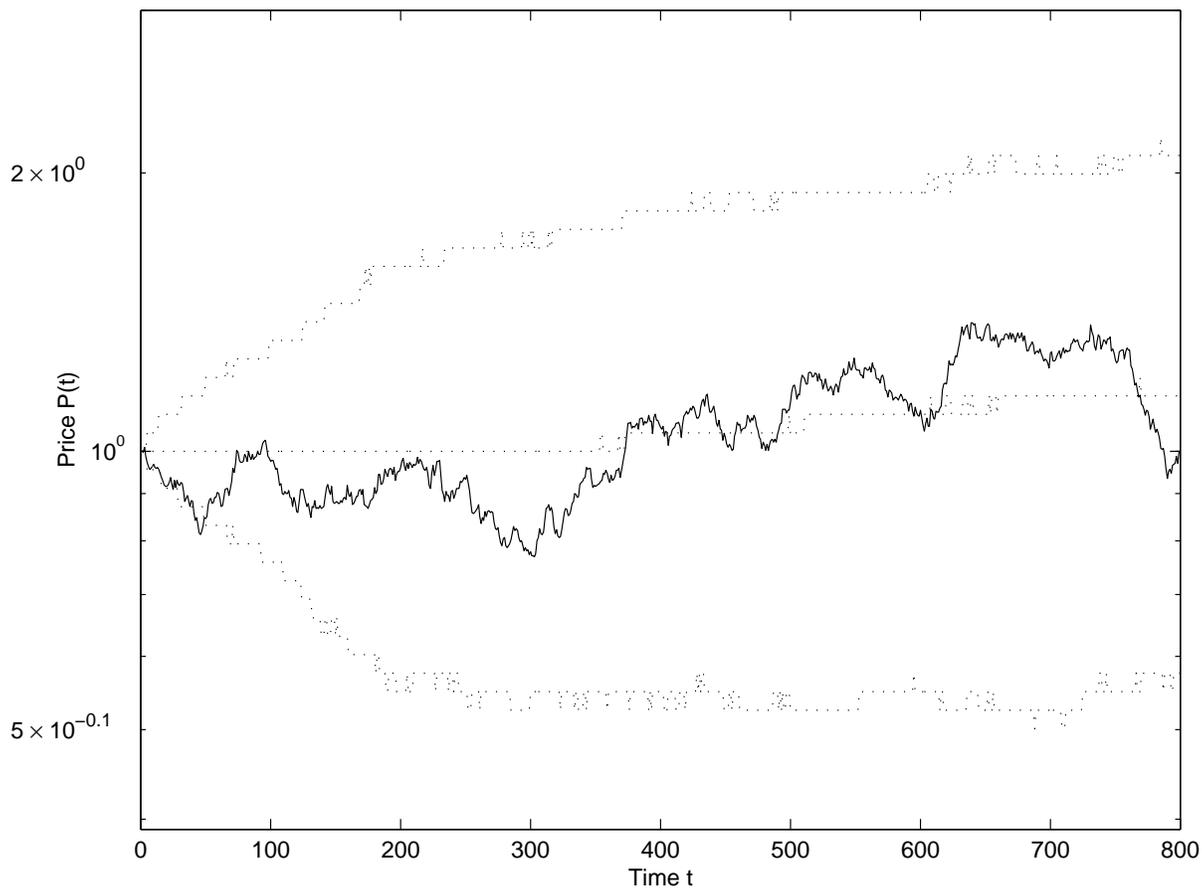}
\caption{\protect\label{Fig4} 
Fat solid line: price $P(t)$ from one configuration of the $\$$-game
with $N=20$ agents, $r=10\%$, $d_0=8\%$ (assuming dividend
$d(t)=d_0 P(t)$). The parameter values used were $s=4, 
m=8, C_0=50$. Stochasticity was introduced via $N$ 
additional ``noise'' agents. The fraction of agents allowed 
to take short positions, $\rho =0$. 
Liquidity parameter $\lambda=0.0025$. 
Thin dotted lines represent the 5\%, 
50\% and 95\% quantiles (from bottom to top) respectively, i.e. at every 
time $t$ out of 
the 1000 different initial configurations only 50 got below the 5\% 
quantile line and similarly for the other quantile lines. 
}
\end{figure}

\begin{figure}[h]
\includegraphics[width=16cm]{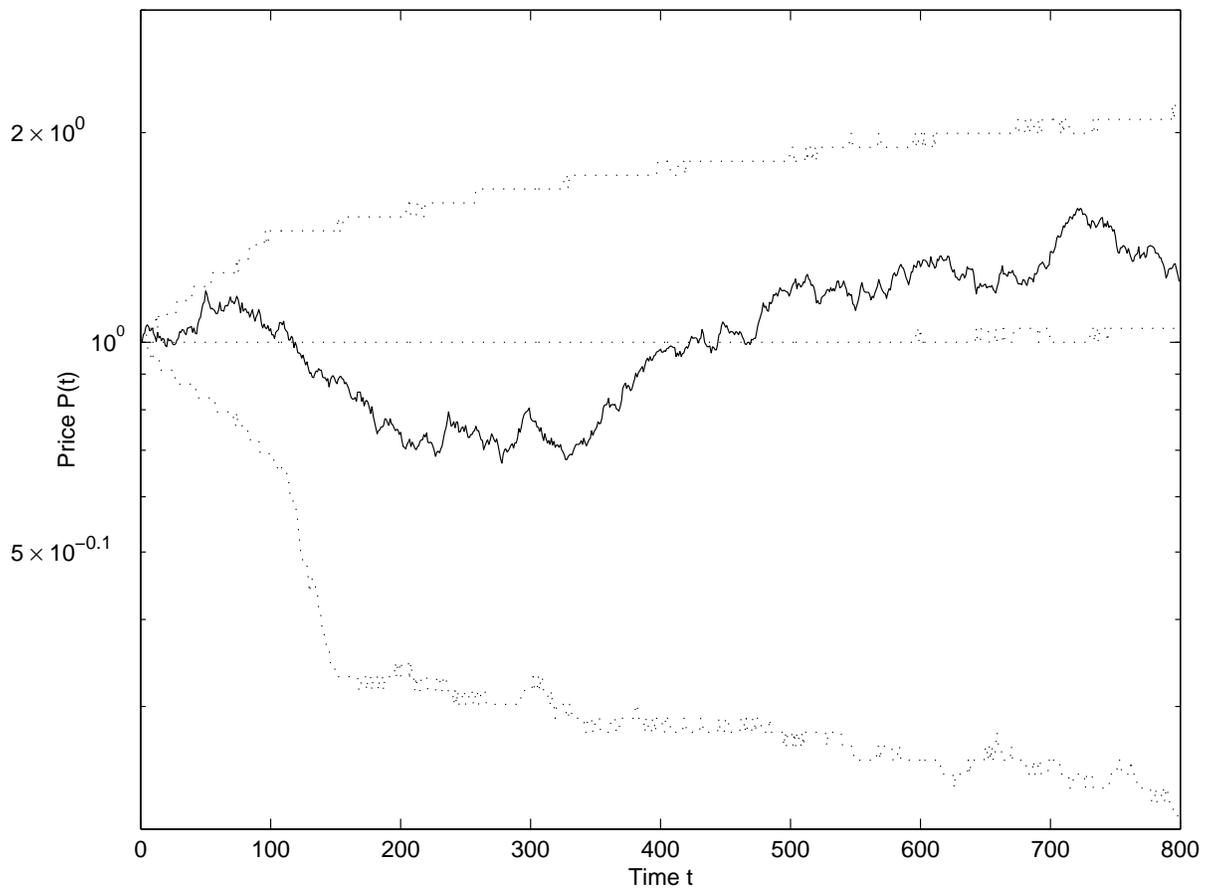}
\caption{\protect\label{Fig5} 
Same as figur~4 but with $\rho=0.2$
}
\end{figure}

\begin{figure}[h]
\includegraphics[width=16cm]{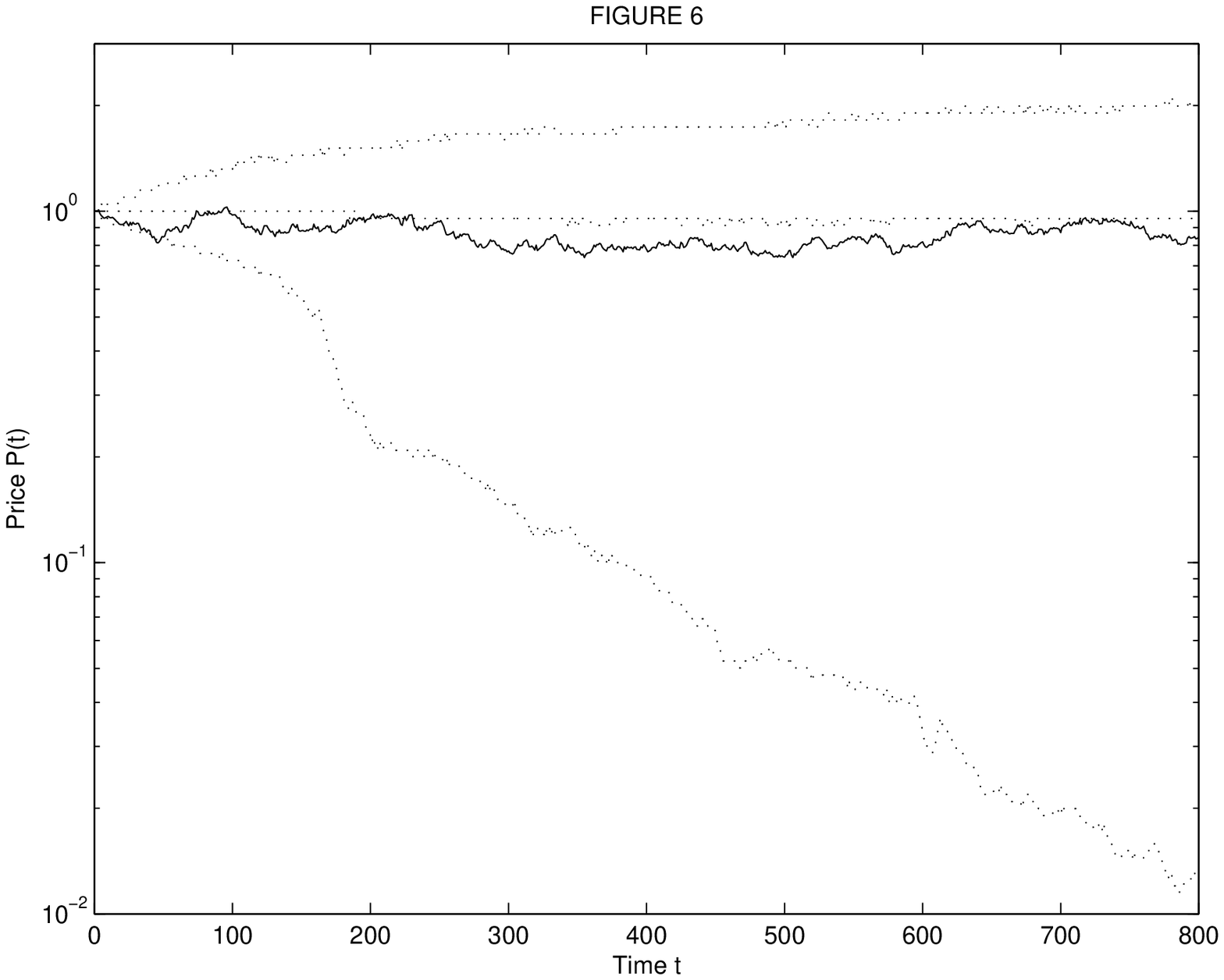}
\caption{\protect\label{Fig6} 
Same as figur~4 but with $\rho=0.4$
}
\end{figure}

\begin{figure}[h]
\includegraphics[width=16cm]{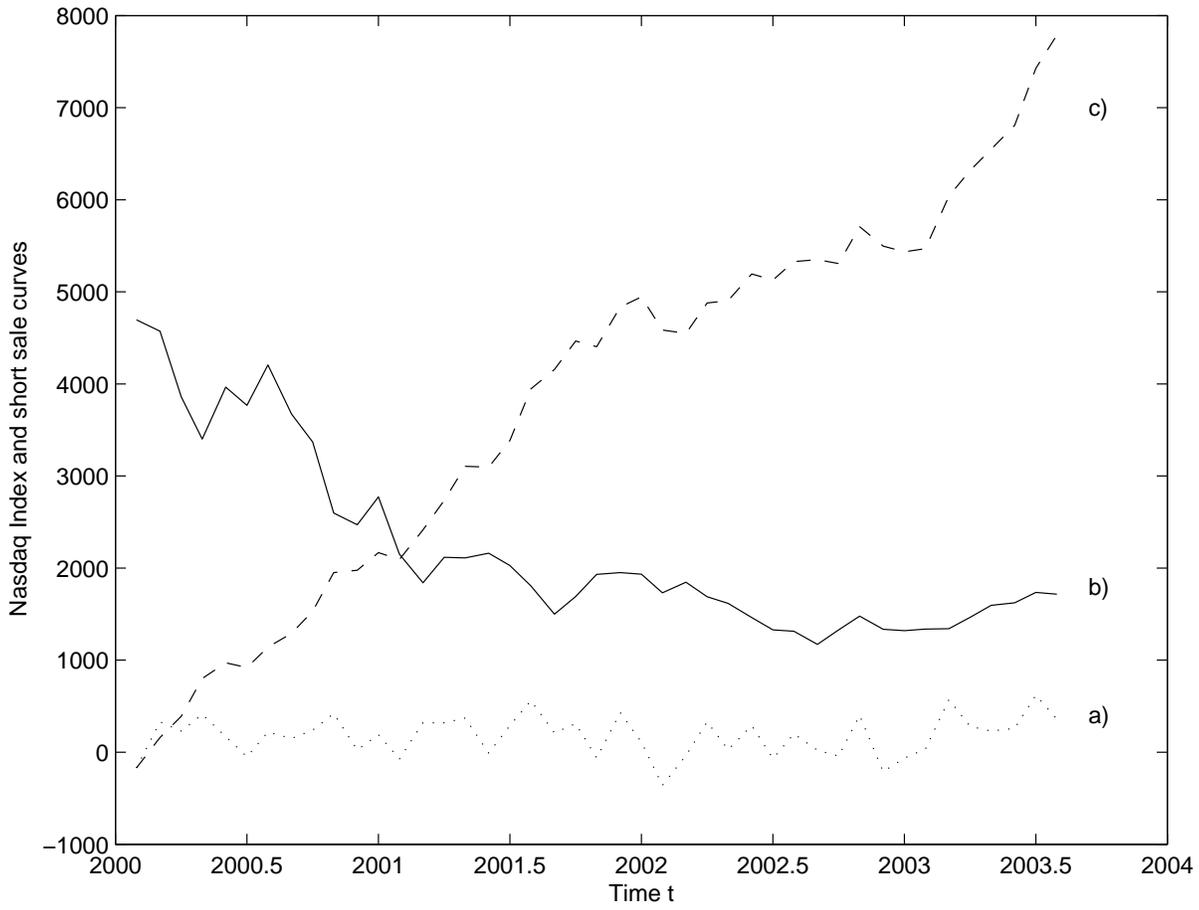}
\caption{\protect\label{Fig7} 
Fat solid line represents the Nasdaq Composite index.
Dotted line represents the fraction 
of change in shares shorted, $n_s(t)$, defined 
as the monthly total number of 
change in shares shorted divided by the total monthly 
share volume. The dashed line represents the cumulative fraction 
of change in shares shorted, $Q_s(t) = \sum_0^t n_s(t)$. 
Both $n_s(t)$ and $Q_s(t)$ have been multiplied with a factor 
$10^5$.
}
\end{figure}


\begin{thebibliography}{}
\bibitem{note1} 
By a market share is understood a portfolio of stocks from 
which the financial market index is composed.
\bibitem{note5} 
A similar set of eq.'s [1-5] were used in \cite{Giardina2} to 
explain bubble formation in an agent based model.
\bibitem{note6} 
Typical values of $(\alpha,r,d)$ will be used rather than 
trying to model a specific market and time period.
\bibitem{note2} 
For simplicity the possibility of leverage is ignored.
\end{thebibliography}
\end{document}